# Analysis of the Pressure - Velocity boundary conditions for the projection method solution of the incompressible Navier-Stokes Equations

**Leonid Pekker**[1)]
FUJIFILM Dimatix Inc., Lebanon NH 03766, USA
Leonid.Pekker@fujifilm.com

**David Pekker**
University of Pittsburgh, Pittsburgh PA 15260, USA
E-mail: pekkerd@pitt.edu

[1)]Author to whom correspondence should be addressed.

## Abstract

The projection method is the standard approach for numerically integrating the incompressible Navier-Stokes equation initial-boundary-value problem. Typical boundary conditions specify either the velocity or the gradient velocity on the boundary. Here, we consider the pressure-tangential-velocity boundary condition in which the tangential components of the velocity and the pressure at the boundary are specified. We (1) show how these boundary conditions can be incorporated into the projection method for use with popular CFD codes like Gerris/Basilisk; (2) perform normal mode analysis to show that the boundary conditions are second-order accurate in time for the standard $\nabla P^{n+1/2}$ projection method and are first-order for the $\nabla P^{n}$ and $\nabla P^{n+1}$ projection methods.

Key Words: incompressible flow, projection method, boundary conditions, Navier Stokes equations, normal mode analysis

## I. Introduction

The incompressible Navier-Stokes equations for a bounded domain Ω are:

$$\frac{\partial \boldsymbol{u}}{\partial t} + (\boldsymbol{u} \cdot \boldsymbol{\nabla})\boldsymbol{u} + \boldsymbol{\nabla} p - \nu \boldsymbol{\nabla} \cdot (\boldsymbol{\nabla} \boldsymbol{u}) = 0, \quad (1)$$

$$\boldsymbol{\nabla} \cdot \boldsymbol{u} = 0, \quad (2)$$

where $\boldsymbol{u}$ is the fluid velocity, $p$ is the pressure of the fluid divided by the mass density of the fluid, and $\nu$ is the kinematic viscosity of the fluid; $\nu$ is assumed to be constants. In this article, we examine the accuracy of different projection methods to the initial-boundary-value problem of Eqs. (1) and (2) with the pressure-tangential-velocity type boundary conditions in the form:

$$p|_{\partial\Omega} = p^b(t), \quad (3a)$$

$$\boldsymbol{u}_\tau|_{\partial\Omega} = \boldsymbol{u}_\tau^b(t), \quad (3b)$$

where $\partial \boldsymbol{u}_n / \partial n|_{\partial\Omega}$ is calculated as

$$\partial \boldsymbol{u}_n / \partial \boldsymbol{n}|_{\partial\Omega} = -\partial \boldsymbol{u}_\tau^b / \partial \boldsymbol{\tau}. \quad (3c)$$

Here, index $b$ corresponds to the boundary; $\partial\Omega$ is the boundary of domain Ω; $\boldsymbol{u}_\tau$ is the tangential velocity and $\boldsymbol{u}_n$ is the normal velocity of the liquid to $\partial\Omega$; and Eq. (3c) is obtained from Eq. (2).

In Section II, we apply boundary conditions (3) for Eqs. (1) and (2) for different projection methods, and then, in Section III, examine them in terms of order accuracy in time by Normal Mode Analysis [1-7]. Concluding remarks are presented in Section IV.

## II. Projection Method

In this section, we consider the standard second-order in space, time-discrete, semi-implicit form [8] of Eqs. (1) and (2) with boundary conditions (3),

$$\frac{\boldsymbol{u}^{n+1} - \boldsymbol{u}^n}{\Delta t} + \boldsymbol{\nabla} p^{n+1/2} = -[(\boldsymbol{u} \cdot \boldsymbol{\nabla})\boldsymbol{u}]^{n+1/2} + \frac{\nu}{2} \nabla^2 (\boldsymbol{u}^{n+1} + \boldsymbol{u}^n), \quad (4)$$

$$\boldsymbol{\nabla} \cdot \boldsymbol{u}^{n+1} = 0, \quad (5)$$

where the upper index corresponds to the number of time steps. In Eqs. (4) and (5), $\boldsymbol{u}^k$ is approximation to $\boldsymbol{u}(k\,\Delta t)$ where $\Delta t$ is the time step; and $[(\boldsymbol{u}\cdot\boldsymbol{\nabla})\boldsymbol{u}]^{n+1/2}$ and $p^{n+1/2}$ are second-order approximations correspondingly to the convective derivative term and pressure at time $(n+1/2)\Delta t$. Here, we assume that $[(\boldsymbol{u}\cdot\boldsymbol{\nabla})\boldsymbol{u}]^{n+1/2}$ is calculated explicitly using the values of this term calculated from previous time steps. Let us solve Eqs. (4) and (5) with boundary condition (3) by a two-step projection method.

***Step*** 1: Solve for the intermediate field $\boldsymbol{u}^*$.

Following the projection method, we introduce an intermediate field $\boldsymbol{u}^*$ that solves

$$\frac{\boldsymbol{u}^*-\boldsymbol{u}^n}{\Delta t}+\boldsymbol{\nabla}p^{n-1/2}=-[(\boldsymbol{u}\cdot\boldsymbol{\nabla})\boldsymbol{u}]^{n+1/2}+\frac{\nu}{2}\nabla^2(\boldsymbol{u}^*+\boldsymbol{u}^n) \qquad \text{(6a)}$$

with boundary conditions

$$\boldsymbol{u}_\tau^*|_{\partial\Omega}=\boldsymbol{u}_\tau^{b,n+1} \;\text{ and }\; \partial\boldsymbol{u}_n^*/\partial\boldsymbol{n}|_{\partial\Omega}=-\partial\boldsymbol{u}_\tau^{b,n+1}/\partial\boldsymbol{\tau}. \qquad \text{(6b)}$$

As follows from Eq. (5),

$$(\boldsymbol{\nabla}\cdot\boldsymbol{u}^*)|_{\partial\Omega}=0 \qquad \text{(6c)}$$

Eqs. (6a) can be written in the form of a modified Poisson equation,

$$\frac{\nu\Delta t}{2}\nabla^2(\boldsymbol{u}^*)-\boldsymbol{u}^*=\Delta t[(\boldsymbol{u}\cdot\nabla)\boldsymbol{u}]^{n+1/2}+\Delta t\nabla p^{n-1/2}-\boldsymbol{u}^n, \qquad \text{(6d)}$$

where the right-hand side of this equation is a source term.

***Step*** 2: Solve for the fields $\boldsymbol{u}^{n+1}$ and $p^{n+1/2}$.

Let us write $\boldsymbol{u}^{n+1}$ as

$$\boldsymbol{u}^{n+1}=\boldsymbol{u}^*-\Delta t\boldsymbol{\nabla}\Phi. \qquad \text{(7a)}$$

Thus, if the scalar function $\Phi$ is known, we can calculate $\boldsymbol{u}^{n+1}$. Taking the divergence of Eq. (7a) and using Eq. (5) we obtain an equation for $\Phi$:

$$\nabla^2\Phi=\frac{1}{\Delta t}(\boldsymbol{\nabla}\cdot\boldsymbol{u}^*), \qquad \text{(7b)}$$

which is a Poisson equation with a source term. Because of $(\boldsymbol{\nabla}\cdot\boldsymbol{u}^*)|_{\partial\Omega}=0$, Eq. (6c), we obtain from Eq. (7b)

$$\nabla^2\Phi|_{\partial\Omega} = 0. \tag{7c}$$

Let us obtain boundary conditions for Eq.(7b). Substituting Eq. (7a) into Eq. (4) we obtain

$$\frac{\boldsymbol{u}^{*}-\Delta t\boldsymbol{\nabla}\Phi-\boldsymbol{u}^{n}}{\Delta t} + \nabla p^{n+1/2} = -[(\boldsymbol{u}\cdot\boldsymbol{\nabla})\boldsymbol{u}]^{n+1/2} + \frac{\nu}{2}\nabla^2(\boldsymbol{u}^{*} - \Delta t\boldsymbol{\nabla}\Phi + \boldsymbol{u}^{n}). \tag{7d}$$

Subtracting Eq. (6d) from Eq. (7d), we obtain

$$-\boldsymbol{\nabla}\Phi + \boldsymbol{\nabla}p^{n+1/2} - \boldsymbol{\nabla}p^{n-1/2} = -\frac{\nu\Delta t}{2}\nabla^2(\boldsymbol{\nabla}\Phi) \rightarrow$$

$$\rightarrow\ p^{n+1/2} - p^{n-1/2} = \Phi - \frac{\nu\Delta t}{2}\nabla^2\Phi. \tag{7e}$$

In Eq. (7e), we put the integration constant equal to zero. Taking into account Eq. (7c), we obtain from Eq. (7e) the following boundary condition for Eq. (7b):

$$\Phi|_{\partial\Omega} = p^{n+1/2}\big|_{\partial\Omega} - p^{n-1/2}\big|_{\partial\Omega}, \tag{7f}$$

where $\Phi|_{\partial\Omega}$ is calculated using the pressure boundary conditions, Eq. (3a). Thus, solving Eq. (7b) with boundary condition (7f), we obtain $\Phi$. Substituting $\Phi$ into Eq. (7a) and in (7e), we calculate fields $\boldsymbol{u}^{n+1}$ and $p^{n+1/2}$, respectively.

It should be noted that Eqs. (1) and (2) with boundary conditions (3) can be numerically solved by the two-step projection method applied to the following digitizing forms of Eqs. (1) and (2), into

$$\frac{\boldsymbol{u}^{n+1}-\boldsymbol{u}^{n}}{\Delta t} + \nabla p^{n} = -[(\boldsymbol{u}\cdot\nabla)\boldsymbol{u}]^{n+1/2} + \frac{\nu}{2}\nabla^2(\boldsymbol{u}^{n+1} + \boldsymbol{u}^{n}), \tag{8a}$$

$$\nabla\cdot\boldsymbol{u}^{n+1} = 0, \tag{8b}$$

and

$$\frac{\boldsymbol{u}^{n+1}-\boldsymbol{u}^{n}}{\Delta t} + \nabla p^{n+1} = -[(\boldsymbol{u}\cdot\nabla)\boldsymbol{u}]^{n+1/2} + \frac{\nu}{2}\nabla^2(\boldsymbol{u}^{n+1} + \boldsymbol{u}^{n}), \tag{9a}$$

$$\nabla\cdot\boldsymbol{u}^{n+1} = 0, \tag{9b}$$

as well. In Eq. (8a), we have used $\nabla p^{n}$ instead of $\nabla p^{n+1/2}$ in Eq. (2) and, in Eq. (9a), $\nabla p^{n+1}$. For the projection method, we need only, in Eq. (6a), to use the gradient of pressure term at the previous time step: $n-1$ for the case of Eqs. (8), and $n$ for the case of Eqs. (9):

$$\frac{\boldsymbol{u}^{*}-\boldsymbol{u}^{n}}{\Delta t}+\nabla p^{n-1}=-[(\boldsymbol{u}\cdot\nabla)\boldsymbol{u}]^{n+1/2}+\frac{\nu}{2}\nabla^{2}(\boldsymbol{u}^{*}+\boldsymbol{u}^{n}), \tag{10a}$$

$$\frac{\boldsymbol{u}^{*}-\boldsymbol{u}^{n}}{\Delta t}+\nabla p^{n}=-[(\boldsymbol{u}\cdot\nabla)\boldsymbol{u}]^{n+1/2}+\frac{\nu}{2}\nabla^{2}(\boldsymbol{u}^{*}+\boldsymbol{u}^{n}). \tag{10b}$$

## III. Normal Mode Analysis of the Pressure Boundary Conditions

In this section, we use the Normal Mode Analysis [1-7] to analyze the time-order of accuracy of the pressure boundary conditions (3) for 2-dimencional incompressible Navier-Stokes. According to the Normal Mode Analysis, we drop the advective derivative term $(\boldsymbol{u}\cdot\nabla)\boldsymbol{u}$ in Eq. (1) reducing Eqs. (1) to a linear Stokes equation, and, as in [5, 6], consider periodic semi-infinite strip $\Omega=[0,\infty)\times[-a,a]$ for $t\geq 0$, Fig. 1:

$$\frac{\partial u}{\partial t}=-\frac{\partial p}{\partial x}+\nu\left(\frac{\partial^{2}u}{\partial x^{2}}+\frac{\partial^{2}u}{\partial y^{2}}\right), \tag{11}$$

$$\frac{\partial v}{\partial t}=-\frac{\partial p}{\partial y}+\nu\left(\frac{\partial^{2}v}{\partial x^{2}}+\frac{\partial^{2}v}{\partial y^{2}}\right), \tag{12}$$

$$\frac{\partial u}{\partial x}+\frac{\partial v}{\partial y}=0, \tag{13}$$

$$u|_{y=a}=u|_{y=-a},\quad v|_{y=a}=v|_{y=-a},\quad p|_{y=a}=p|_{y=-a}, \tag{14}$$

$$u|_{x=\infty}=0,\quad v|_{x=\infty}=0,\quad p|_{x=\infty}=0, \tag{15}$$

$$p|_{x=0}=p_{0}(y,t),\quad v|_{x=0}=v_{0}(y,t),\quad \left.\frac{\partial u}{\partial x}\right|_{x=0}=-\frac{\partial v_{0}(y,t)}{\partial y}, \tag{16}$$

where $p_0$ and $v_0$ are the driving pressure and the driving fluid tangential velocity, respectively, at $x=0$. Here, boundary conditions (14) correspond to periodicity of solution along the y-axis; boundary conditions (15) correspond to decaying the flow at $x\to\infty$, and boundary conditions (16) correspond to pressure-tangential-velocity boundary condition (3) at $x=0$.

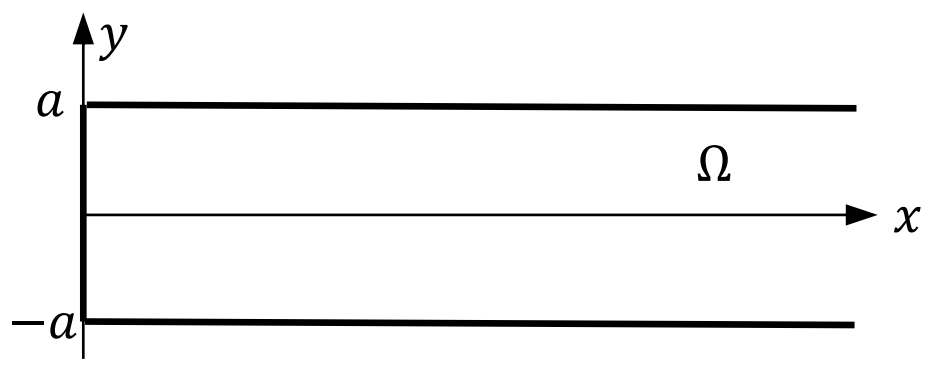

Fig. 1. Domain for the normal model analysis.

Applying the Fourier transformation in $y$ with wavenumber $k = 2\pi m/a$ where $m = 1,2, \ldots$ (to satisfy boundary conditions (14)) and the Fourier transformation in $t$ to Eqs. (11) – (13), and assuming

$$p_0(y,t) = p_0 e^{i\omega t + iky} \text{ and } v_0(y,t) = v_0 e^{i\omega t + iky}, \quad (17)$$

$$p(x,y,t) = P(x) e^{i\omega t + iky},\ u(x,y,t) = U(x) e^{i\omega t + iky}, \text{ and } v(x,y,t) = V(x) e^{i\omega t + iky}, \quad (18)$$

Eqs. (11) – (13), (15), (16) reduce to the following view

$$i\omega U = -\frac{dP}{dx} + \nu\left(\frac{d^2U}{dx^2} - k^2 U\right), \quad (19)$$

$$i\omega V = -ikP + \nu\left(\frac{d^2V}{dx^2} - k^2 V\right), \quad (20)$$

$$ikV + \frac{dU}{dx} = 0, \quad (21)$$

$$U(\infty) = 0,\ \ V(\infty) = 0,\ \ P(\infty) = 0, \quad (22)$$

$$P|_{x=0} = p_0,\ \ U|_{x=0} = -ikv_0,\ \ V|_{x=0} = v_0. \quad (23)$$

Taking derivative from Eq. (19) and multiplying Eq. (20) by $ik$ and then summing the obtained equations together and then using Eq. (21), we obtain equation for $P$,

$$\left(\frac{d^2}{dx^2} - k^2\right) P = 0. \quad (24)$$

Solving Eq. (23) for $P$ with boundary condition (23) we obtain

$$P = p_0 \exp(-kx). \quad (25)$$

Substituting Eq. (25) into Eqs. (19) and (20) we obtain a set of the ordinary differential equations for $U(x)$ and $V(x)$:

$$\frac{d^2U}{dx^2} = \left(k^2 + \frac{i\omega}{\nu}\right) U - \frac{k}{\nu} p_0 \mathrm{e}^{-kx}, \quad (26)$$

$$\frac{d^2V}{dx^2} = \left(k^2 + \frac{i\omega}{\nu}\right) V + \frac{ik}{\nu} p_0 \mathrm{e}^{-kx}. \quad (27)$$

Solving this set of equations with boundary conditions (22) and (23) we obtain:

$$U = \left(\frac{ikv_0}{\left(k^2 + \frac{i\omega}{\nu}\right)^{1/2}} + \frac{ik^2 p_0}{\omega\left(k^2 + \frac{i\omega}{\nu}\right)^{1/2}}\right) exp\left(-\left(k^2 + \frac{i\omega}{\nu}\right)^{1/2} x\right) - \frac{ikp_0}{\omega} exp(-kx), \quad (28)$$

$$V = \left(v_0 + \frac{kp_0}{\omega}\right) exp\left(-\left(k^2 + \frac{i\omega}{\nu}\right)^{1/2} x\right) - \frac{kp_0}{\omega} exp(-kx). \quad (29)$$

Thus, the set of Eqs. (25), (28) - (29) solves the system of Eqs. (19) – (23).

Following the Normal Mode Analysis, we extend the form Eq. (4) onto Eqs. (11) - (16):

$$\frac{u^{n+1}-u^n}{\Delta t}+\frac{\partial p^{n+1/2}}{\partial x}=\frac{\nu}{2}\left(\frac{\partial^2}{\partial x^2}+\frac{\partial^2}{\partial y^2}\right)(u^{n+1}+u^n), \tag{30}$$

$$\frac{v^{n+1}-v^n}{\Delta t}+\frac{\partial p^{n+1/2}}{\partial y}=\frac{\nu}{2}\left(\frac{\partial^2}{\partial x^2}+\frac{\partial^2}{\partial y^2}\right)(u^{n+1}+u^n). \tag{31}$$

$$\frac{\partial u^{n+1}}{\partial x}+\frac{\partial v^{n+1}}{\partial x}=0. \tag{32}$$

$$u^{n+1}|_{y=a}=u^{n+1}|_{y=-a},\quad v^{n+1}|_{y=a}=v^{n+1}|_{y=-a},\quad p^{n+1/2}\big|_{y=a}=p^{n+1/2}\big|_{y=-a}, \tag{33}$$

$$u^{n+1}|_{x=\infty}=0,\quad v^{n+1}|_{x=\infty}=0,\quad p^{n+1/2}\big|_{x=\infty}=0, \tag{34}$$

$$\left.\frac{\partial u^{n+1}}{\partial y}\right|_{x=0}=-\frac{\partial v_0(y,\Delta t(n+1))}{\partial y},\quad v^{n+1}|_{x=0}=v_0(y,\Delta t(n+1)),\quad p^{n+1/2}\big|_{x=0}=p_0(y,\Delta t(n+1/2)). \tag{35}$$

Applying the Fourier transformation in $y$ with wavenumber $k=2\pi m/a$ where $m=1,2,\dots$ and the Fourier transformation in $t$ to Eqs. (28) – (32), and taking into account that

$$p_0^{n+1/2}=p_0e^{i\omega(n+1/2)\Delta t+iky}\quad\text{and}\quad v_0^n=v_0e^{i\omega n\Delta t+iky}, \tag{36}$$

$$p^{n+1/2}=P(x)e^{i\omega(n+1/2)\Delta t+iky},\ u^n=U(x)e^{i\omega n\Delta t+iky},\ v^n=V(x)e^{i\omega n\Delta t+iky}, \tag{37}$$

we obtain a time-discrete form of Eqs. (19) – (23):

$$\frac{U}{\Delta t}\left(e^{i\omega\Delta t}-1\right)+\frac{dP}{dx}e^{i0.5\omega\Delta t}=\frac{\nu}{2}\left(1+e^{i\omega\Delta t}\right)\left(\frac{d^2}{dx^2}-k^2\right)U, \tag{38}$$

$$\frac{V}{\Delta t}\left(e^{i\omega\Delta t}-1\right)+ikPe^{i0.5\omega\Delta t}=\frac{\nu}{2}\left(1+e^{i\omega\Delta t}\right)\left(\frac{d^2}{dx^2}-k^2\right)V, \tag{39}$$

$$\frac{dU}{dx}+ikV=0, \tag{40}$$

$$V|_{x=\infty}=0,\quad U|_{x=\infty}=0,\quad P|_{x=\infty}=0, \tag{41}$$

$$V|_{x=0}=v_0,\quad \left.\frac{dU}{dx}\right|_{x=0}=-ikv_0,\quad P|_{x=0}=p_0. \tag{42}$$

Taking derivative $dx$ from Eq. (38) and multiplying Eq. (39) by $ik$ and summing the obtained equations, and then using Eq. (40) we obtain the following equation for $P$:

$$\frac{d^2P}{dx^2}-k^2P=0. \tag{43}$$

Solving Eq. (43) for $P$ with boundary conditions (41) and (42), we obtain

$$P(x) = p_0 e^{-kx}. \tag{44}$$

Substituting Eq. (44) into Eqs. (38) and (39), these equations can be reduced to the following view:

$$\frac{d^2U}{dx^2} = \left(k^2 + \frac{2(e^{i\omega\Delta t}-1)}{\nu\Delta t(1+e^{i\omega\Delta t})}\right)U - p_0\frac{k}{\nu}\cdot\frac{2e^{i0.5\omega\Delta t}}{(1+e^{i\omega\Delta t})}e^{-kx}, \tag{45}$$

$$\frac{d^2V}{dx^2} = \left(k^2 + \frac{2(e^{i\omega\Delta t}-1)}{\nu\Delta t(1+e^{i\omega\Delta t})}\right)V + p_0\frac{ik}{\nu}\frac{2e^{i0.5\omega\Delta t}}{(1+e^{i\omega\Delta t})}e^{-kx}. \tag{46}$$

A solution of this set of equations with boundary conditions (41) and (42) are:

$$U = \left(\frac{ikv_0}{\left(k^2+\frac{i\omega\delta}{\nu}\right)^{1/2}} + \frac{ik^2p_0\xi}{\omega\left(k^2+\frac{i\omega\delta}{\nu}\right)^{1/2}}\right)exp\left(-\left(k^2+\frac{i\omega\delta}{\nu}\right)^{1/2}x\right) - \frac{ikp_0\xi}{\omega}exp(-kx), \tag{47}$$

$$V = \left(v_0 + \frac{kp_0\xi}{\omega}\right)exp\left(-\left(k^2+\frac{i\omega\delta}{\nu}\right)^{1/2}x\right) - \frac{kp_0\xi}{\omega}exp(-kx), \tag{48}$$

$$\xi = \frac{i\Delta t\omega e^{i\omega\Delta t/2}}{e^{i\omega\Delta t}-1}, \quad \delta = \frac{2(e^{i\omega\Delta t}-1)}{i\Delta t\omega(e^{i\omega\Delta t}+1)}. \tag{49}$$

Comparing Eq. (28) with Eq. (47) and Eq. (29) with Eq. (48), we see that they match each-other where $\xi = \delta = 1$; the equations for pressure, Eqs. (25) and (44), are identical. Taking into account that the period of oscillation, $2\pi/\omega$, should be much larger than integration time step $\Delta t$, we obtain that

$$\xi|_{\Delta t\omega\ll 1} = 1 + \frac{(\omega\Delta t)^2}{24}, \qquad \delta|_{\Delta t\omega\ll 1} = 1 + \frac{(\omega\Delta t)^2}{12}. \tag{50}$$

Thus, we have shown that solving Eqs. (1) - (2) with boundary condition (3) by the projection method applied to Eq. (4) - (5) with boundary conditions (3) provides the accuracy of the second-order in time.

Applying the Normal Mode Analysis to Eqs. (8) and (9) with boundary conditions (3) shows (see Appendix) that, for both cases, the projection method provides the accuracy of the first order in time: that, when $\Delta t\omega \ll 1$, $\xi = 1 - i\omega\Delta t/2$ and $\xi = 1 + i\omega\Delta t/2$ for Eqs. (8) and Eqs. (9) respectively; the formulas for $\delta$ are the same for all three cases and given by Eq. (49) and, in case of $\Delta t\omega \ll 1$, by Eq. (50).

## IV. Concluding Remarks

In this article, we suggested the two-step projection methods for solving the incompressible Navier-Stokes equations with the pressure-velocity boundary conditions, in which the pressure and the tangential component of the velocity are given on a boundary, and the gradient of the normal component of the velocity at this boundary is calculated using the volume conservation law. Using the Normal Mode Analysis, we show the standard $\nabla P^{n+1/2}$ projection method provides the second-order accuracy in time, and $\nabla P^{n}$ and $\nabla P^{n+1}$ projection methods provide the first-order accuracy.

These projection methods can be easily incorporated in popular CFD codes like Gerris/Basilisk.

## Acknowledgements

The authors would like to express their sincere gratitude to Dan Barnett for his kind support of this project, and to Sean Cahill and Robert Vivanco for their very valuable comments. The authors would like also to acknowledge the support of FujiFilm Dimatix for conducting this research.

## Appendix: Normal Mode Analysis of Eq. (8) and (9) with Boundary Conditions (3)

Applying Normal Mode Analysis to the set of Eqs. (8) and to the set of Eqs. (9) reduce them to the following forms:

$$\frac{u^{n+1}-u^{n}}{\Delta t}+\frac{\partial p^{n}}{\partial x}=\frac{\nu}{2}\left(\frac{\partial^2}{\partial x^2}+\frac{\partial^2}{\partial y^2}\right)(u^{n+1}+u^{n}), \quad \text{(A1)}$$

$$\frac{v^{n+1}-v^{n}}{\Delta t}+\frac{\partial p^{n}}{\partial y}=\frac{\nu}{2}\left(\frac{\partial^2}{\partial x^2}+\frac{\partial^2}{\partial y^2}\right)(u^{n+1}+u^{n}), \quad \text{(A2)}$$

$$\frac{\partial u^{n+1}}{\partial x}+\frac{\partial v^{n+1}}{\partial x}=0, \quad \text{(A3)}$$

and

$$\frac{u^{n+1}-u^{n}}{\Delta t}+\frac{\partial p^{n+1}}{\partial x}=\frac{\nu}{2}\left(\frac{\partial^2}{\partial x^2}+\frac{\partial^2}{\partial y^2}\right)(u^{n+1}+u^{n}), \quad \text{(B1)}$$

$$\frac{v^{n+1}-v^{n}}{\Delta t}+\frac{\partial p^{n+1}}{\partial y}=\frac{\nu}{2}\left(\frac{\partial^2}{\partial x^2}+\frac{\partial^2}{\partial y^2}\right)(u^{n+1}+u^{n}), \quad \text{(B2)}$$

$$\frac{\partial u^{n+1}}{\partial x}+\frac{\partial v^{n+1}}{\partial x}=0. \quad \text{(B3)}$$

Here and below equations starting with “A” correspond to the case of $\nabla p^{n}$ (Eqs. 8), and equations starting with “B” to the case of $\nabla p^{n+1}$ (Eqs. 9). The boundary conditions for $u^{n+1}$ and $v^{n+1}$ are the same for both set of equations and given by Eqs. (33) and (34). In formulas for pressure boundary conditions, in Eqs. (33) and (33), we must use indexes $n$ and $n+1$ correspondingly for case A and case B.

As in Section III, applying the Fourier transformation in $y$ with wavenumber $k = 2\pi m/a$ where $m = 1,2, ...$, and the Fourier transformation in $t$ to Eqs. (A1) – A(3), and to Eqs. (B1) – (B3), and to the modified boundary conditions (33) – (33), and assuming

$$p_0^n = p_0 e^{i\omega n\Delta t + iky} \quad \text{and} \quad v_0^n = v_0 e^{i\omega n\Delta t + iky}, \tag{A4}$$

$$p^n = P(x) e^{i\omega n\Delta t + iky}, \;\; u^n = U(x) e^{i\omega n\Delta t + iky}, \;\; v^n = V(x) e^{i\omega n\Delta t + iky}, \tag{A5}$$

and

$$p_0^{n+1} = p_0 e^{i\omega (n+1)\Delta t + iky} \quad \text{and} \quad v_0^n = v_0 e^{i\omega n\Delta t + iky}, \tag{B4}$$

$$p^{n+1} = P(x) e^{i\omega (n+1)\Delta t + iky}, \;\; u^n = U(x) e^{i\omega n\Delta t + iky}, \;\; v^n = V(x) e^{i\omega n\Delta t + iky}, \tag{B5}$$

we obtain the digitized forms of Eqs. (19) – (23):

$$\frac{U}{\Delta t}\left(e^{i\omega\Delta t} - 1\right) + \frac{dP}{dx} = \frac{\nu}{2}\left(1 + e^{i\omega\Delta t}\right)\left(\frac{d^2}{dx^2} - k^2\right)U, \tag{A6}$$

$$\frac{V}{\Delta t}\left(e^{i\omega\Delta t} - 1\right) + ikP = \frac{\nu}{2}\left(1 + e^{i\omega\Delta t}\right)\left(\frac{d^2}{dx^2} - k^2\right)V, \tag{A7}$$

$$\frac{dU}{dx} + ikV = 0, \tag{A8}$$

$$V|_{x=\infty} = 0, \quad U|_{x=\infty} = 0, \;\; P|_{x=\infty} = 0, \tag{A9}$$

$$V|_{x=0} = v_0, \quad \left.\frac{dU}{dx}\right|_{x=0} = -ikv_0, \;\; P|_{x=0} = p_0, \tag{A10}$$

and

$$\frac{U}{\Delta t}\left(e^{i\omega\Delta t} - 1\right) + \frac{dP}{dx} e^{i\omega\Delta t} = \frac{\nu}{2}\left(1 + e^{i\omega\Delta t}\right)\left(\frac{d^2}{dx^2} - k^2\right)U, \tag{B6}$$

$$\frac{V}{\Delta t}\left(e^{i\omega\Delta t} - 1\right) + ikPe^{i\omega\Delta t} = \frac{\nu}{2}\left(1 + e^{i\omega\Delta t}\right)\left(\frac{d^2}{dx^2} - k^2\right)V, \tag{B7}$$

$$\frac{dU}{dx} + ikV = 0, \tag{B8}$$

$$V|_{x=\infty} = 0, \quad U|_{x=\infty} = 0, \;\; P|_{x=\infty} = 0, \tag{B9}$$

$$V|_{x=0} = v_0, \quad \left.\frac{dU}{dx}\right|_{x=0} = -ikv_0, \;\; P|_{x=0} = p_0. \tag{B10}$$

Eliminating pressure terms in the same way as it has been done in Section III, these sets of equations reduce to the following forms:

$$\frac{d^2U}{dx^2} = \left(k^2 + \frac{2(e^{i\omega\Delta t}-1)}{\nu\Delta t(1+e^{i\omega\Delta t})}\right)U - p_0\frac{k}{\nu}\cdot\frac{2}{(1+e^{i\omega\Delta t})}e^{-kx}, \tag{A11}$$

$$\frac{d^2V}{dx^2} = \left(k^2 + \frac{2(e^{i\omega\Delta t}-1)}{\nu\Delta t(1+e^{i\omega\Delta t})}\right)V + p_0\frac{ik}{\nu}\frac{2}{(1+e^{i\omega\Delta t})}e^{-kx}, \tag{A12}$$

and

$$\frac{d^2U}{dx^2} = \left(k^2 + \frac{2(e^{i\omega\Delta t}-1)}{\nu\Delta t(1+e^{i\omega\Delta t})}\right)U - p_0\frac{k}{\nu}\cdot\frac{2e^{i\omega\Delta t}}{(1+e^{i\omega\Delta t})}e^{-kx}, \tag{B11}$$

$$\frac{d^2V}{dx^2} = \left(k^2 + \frac{2(e^{i\omega\Delta t}-1)}{\nu\Delta t(1+e^{i\omega\Delta t})}\right)V + p_0\frac{ik}{\nu}\frac{2e^{i\omega\Delta t}}{(1+e^{i\omega\Delta t})}e^{-kx}. \tag{B12}$$

A Solution of Eqs. (A.11) and (A12) with boundary conditions (A9) and (A10), and Eqs. (B11) and (B12) with boundary conditions (B9) and (B10) are given by Eqs. (47) and (48) with $\delta$ from Eq. (49) and

$$\xi = \frac{i\Delta t\omega}{e^{i\omega\Delta t}-1} \tag{A13}$$

and

$$\xi = \frac{i\Delta t\omega e^{i\omega\Delta t}}{e^{i\omega\Delta t}-1}. \tag{B13}$$

In the case of $\omega\Delta t \ll 1$, Eqs. (A13) and (B13) reduce to:

$$\xi|_{\Delta t\omega\ll 1} = 1 - \frac{i\omega\Delta t}{2}, \tag{A14}$$

$$\xi|_{\Delta t\omega\ll 1} = 1 + \frac{i\omega\Delta t}{2}. \tag{B14}$$

## References

[1] S. A. Orszag, M. Israeli, M. O. Deville, ‘Boundary conditions for incompressible flows’, J. Sci. Comput. 1986; 1:75-111.

[2] G. E. Karniadakis and M. Israeli, ‘High-order splitting methods for the incompressible Navier-Stokes equations’ J. Comput. Physics 1991; 97, 414-443.

[3] W. D. Hensham, H-O. Kreiss, and L. G. M. Reyna, ‘A fourth-order-accurate difference approximation for the incompressible Navier-Stokes equations’, Computers Fluids 1994; 4: 575-593.

[4] W. E and J-G. Liu, ‘Projection method II: Godunov-Ryabenski Analysis’, SIAM J. Numer. Anal. 1996; 33:1597-1621.

[5] J. C. Strikwerda and Y. S. Lee, ‘The accuracy of the fractional step method’, SIAM J. Numer. Anal. 1999; 37: 37-47.

[6] D. L. Brown , R. Cortez, and L. Minion, ‘Accurate Projection Methods for the incompressible Navier-Stokes equations’, J. Comp. Phys. 2001; 268:464-499.

[7] W. E and J-G. Liu, ‘Gauge method for viscous incompressible flows’, Comp. Math. Sc. 2003; 1, 317-332.

[8] J. B. Bell, P. Colella, and H. M. Glaz, ‘A second-order projection method for the incompressible Navier-Stokes equations”, J. Comp. Phys. 1989; 85, 257-283.